\documentclass{PoS}

\def\mpi{M_\pi}
\def\mn{M_{\scriptscriptstyle N}}
\def\etal{{\it et al.\ }}
\def\HBCPT{HB$\chi$PT}

\title{Compton scattering from the proton: An analysis using the delta expansion up to N3LO}

\ShortTitle{Compton scattering from the proton: An analysis using the delta expansion up to N3LO}

\author{\speaker{Judith A. McGovern}\\%
        University of Manchester\\
        E-mail: \email{judith.mcgovern@manchester.ac.uk}}
\author{Harald W. Griesshammer\\
        George Washington University\\
        E-mail: \email{hgrie@gwu.edu}}
\author{Daniel R. Phillips\\
        University of Manchester and Ohio University\thanks{Permanent affiliation}\\
        E-mail: \email{phillips@phy.ohiou.edu}}
\author{Deepshikha Shukla\\
        George Washington University\thanks{Present affiliation: The University of North Carolina at Chapel Hill}\\
        E-mail: \email{choudhur@gwu.edu}}

\abstract{We report on a chiral effective field theory  
calculation of Compton scattering from the proton. Our calculation  
includes pions, nucleons, and the Delta(1232) as explicit degrees of  
freedom. It uses the ``delta expansion'', and so implements the  
hierarchy of scales $\mpi < \Delta\equiv M_\Delta-\mn < \Lambda_\chi$, through a counting $\mpi/ 
\Delta \sim \Delta/\Lambda_\chi \sim \delta$. 
In this expansion the power counting in the vicinity of the Delta peak changes, and resummation of the loop graphs associated with the Delta width is indicated. This is designed to extend the region of $\chi$PT applicability to momenta $p\sim\Delta$.

We have computed the nucleon Compton amplitude in  
the delta expansion up to N$^3$LO---${\cal O}(e^2 \delta^4)$---for photon energies  
$\omega \sim \mpi$. 
This is the first order at which the proton Compton scattering amplitudes receive contributions from contact operators which encode contributions to the spin-independent polarisabilities from states with energies $\sim \Lambda_\chi$.  We fit the coefficients of these two operators to the experimental proton Compton data that
has been taken in the relevant photon-energy domain, and are in a position to extract
new results for $\alpha_p$ and $\beta_p$.}

\FullConference{6th International Workshop on Chiral Dynamics\\
		 July 6-10 2009\\
		 Bern, Switzerland}

\begin{document}

\section{Introduction to Compton scattering and Chiral Dynamics}

Chiral dynamics in the baryonic sector is typically thought of as the study of the interactions of pions and nucleons.  However the dictates of electromagnetic gauge invariance mean that chiral symmetry also strongly constrains the interactions of both with photons, and so Compton scattering from the nucleon is as fundamental a probe of chiral dynamics as pion-nucleon or nucleon-nucleon scattering.  The lowest-order term in the Compton scattering amplitude (the long-wavelength limit) is the Thomson term which is reproduced by $\chi$PT but which, depending as it does only on the nucleon charge and mass, is independent of chiral dynamics. However, at shorter wavelengths the probing proton starts to be sensitive to the structure of the target.  At NLO in heavy baryon chiral perturbation theory (\HBCPT) the dominant new contribution comes from a single pion loop with photons coupling to the pion or to the $\pi$N vertex (see Fig.~\ref{jmcg:fig1}), and hence a prediction can be made for these structure effects.  This includes, but is not limited to,  the numbers known as the polarisabilities of the nucleon; the latter are the first terms in an expansion of the six scattering amplitudes in powers of the photon energy.

The application of chiral dynamics to Compton scattering dates right back to the dawn of baryon $\chi$PT, and most famously the lowest-order predictions for the electric and magnetic polarisability of the proton in \HBCPT, 
$\alpha=12.5\times 10^{-4}$fm${}^3$ and $\beta=1.2\times 10^{-4}$fm${}^3$ \cite{jmcg:ber92} are in extremely good agreement with experimental determinations.  Higher-order calculations \cite{jmcg:ber93,jmcg:ber93a}, and calculations of spin polarisabilities \cite{jmcg:ber95, jmcg:ji00,jmcg:KMB99,jmcg:gellas,jmcg:KMB00} followed.  The first studies to systematically compare the full fourth-order predicted cross-section to a compendium of experimental data were published by some of the current authors \cite{jmcg:McG01,jmcg:bmmpv03}, first with values of $\alpha$ and $\beta$ taken from the Particle Data Group, then using them as fit parameters (at fourth order, they are no longer predicted but have contributions from LECs).  These studies obtained an excellent fit at low energies, and clearly demonstrated the pion-production cusp.  Furthermore the fourth-order results are in markedly better agreement with the data than the third-order ones (even when the freedom to fit the polarisabilities is not exploited). However they also showed that the cross-section completely failed to keep pace with the sharp rise in the experimental data above the cusp, particularly at backward angles.  This is not surprising: from around 200~MeV the experimental cross section is dominated by the Delta resonance.  While contributions from the Delta enter the Delta-less theory via the LECs, these clearly cannot describe the resonance.  Nonetheless, by restricting the fit to appropriate regions of energy and angle, the following  results were obtained \cite{jmcg:bmmpv03}:
\begin{eqnarray*}
\alpha_p&\!=\!&(12.1 \pm 1.1~({\rm stat.}))_{-0.5}^{+0.5}~({\rm theory})\times 10^{-4}~{\rm fm}^3\\
\beta_p&\!=\!&(3.4 \pm 1.1~({\rm stat.}))_{-0.1}^{+0.1}~({\rm theory})\times 10^{-4}~{\rm fm}^3.
\end{eqnarray*}

Work has been done on including the Delta in chiral effective theories from the earliest days of baryon $\chi$PT, motivated partly by its role as the large-$N_C$ degenerate partner of the nucleon, and partly by its experimental relevance.    However there is no universal agreement as to the correct power-counting; it is not obvious how to treat the scale $\Delta=M_\Delta-\mn\approx 2 \mpi$.  The most commonly adopted solution is to treat them as proportional to the same scale (the ``small scale expansion''), a pragmatic treatment which however is not strictly in the spirit of $\chi$PT as $\Delta$ does not vanish in the $m_q\to0$ limit.  With this treatment $\pi\Delta$ loops come in at the same order as $\pi N$ loops. In the particular case of Compton scattering the lowest-order prediction for $\alpha$ and $\beta$ is destroyed with this counting.  More recently Pascalutsa and 
Phillips proposed a more sophisticated treatment in which the power counting differs in different energy regions, depending on which scales are enhanced.  At low energies they introduce the ratio $\delta$, with $\delta\equiv\frac{\mpi}{\Delta}\sim \frac {\Delta}{\Lambda_\chi}$  so that $\delta^2\equiv \left(\frac{\Delta}{\Lambda_\chi}\right)^2 \sim \frac{\mpi}{\Lambda_\chi}$.  With this ``delta-expansion'', the $\pi\Delta$ loops enter at one order higher than $\pi N$ loops.

Compton scattering cross sections calculated at lowest non-trivial order in the small scale expansion were compared to data by Hildebrandt \etal \cite{jmcg:HGHP04}.  Because of the very large contribution of the Delta to 
$\alpha$ and $\beta$, they had to introduce counterterms for these, which they then fit to data in a similar fashion to Ref.~\cite{jmcg:bmmpv03}.  Interestingly, in spite of the rather different input and data set, the results were very similar: 
\begin{displaymath}
\alpha_p =11.5\pm1.4_{\rm stat}\times 10^{-4} \, {\rm fm}^3,\qquad
\beta_p = 3.4\pm1.6_{\rm stat}\times 10^{-4} \,{\rm fm}^3.
\end{displaymath}
However the Delta-full results were able to describe data up to around 200~MeV lab energy, unlike the  Delta-less results.

Clearly the next step is to include both the Delta and higher-order effects, and that is the aim of the current program.  Very recently Lensky and Pascalutsa have also taken steps in this direction, using a relativistic formulation.  However they have not reached the order at which they must fit $\alpha$ and $\beta$, and they have so far only compared to a subset of world data \cite{jmcg:len09}.

\section{Comments on the proton Compton database}

Modern experiments measure the differential cross section for tagged photons over a 
range of energies and angles. 
Between 1956 and 1995, twelve Compton scattering experiments took data at energies below 200~MeV. In a useful paper Baranov {\it et al} \cite{jmcg:BLPS01} examined the 
world data on proton Compton scattering, and demonstrated that the
data from the 50s and 60s was compatible with the more modern data
from 1974 onwards, and was useful in reducing errors. In addition, the
most recent and extensive experiment used the TAPS detector and MAMI tagged photon facility at Mainz \cite{jmcg:Olmos01}.  

As part of the current project we have evaluated the consistency of the various data sets in as model-independent a fashion as we can.  We find that it is crucial to allow the normalizations to float (as done in Ref.~\cite{jmcg:BLPS01,jmcg:bmmpv03} but not in \cite{jmcg:HGHP04}).  For the Mainz data  we increase the statistical error by including a ``random  systematic'' error of 5\% in quadrature, as  described in Ref.~\cite{jmcg:Olmos01}. Some of the very early data cannot be accommodated, nor---as is well known---can the Bonn data of Genzel \etal \cite{jmcg:genz76}.  With these caveats, we find that the world data is reasonably consistent in the range 0--240~MeV.  There is enough noise in the data sets that a $\chi^2$ of one per degree of freedom is not obtainable though. 

Such data has traditionally been analyzed in one of two ways.  Early experiments typically took very low energy data (up to around 100~MeV) and fitted to a low-energy expansion of the cross section which included only Born contributions and the electric and magnetic polarisabilities, with these as fit  parameters.  However at these energies the Born cross section dominates, and the sensitivity to $\alpha$ and $\beta$ is limited.  A desire to use the full range of data up to the region of the pion-production threshold meant that later authors turned to dispersion relations to obtain a prediction for the variation of the cross section with energy, leaving only certain combinations of the polarisabilities as free parameters to be fit.  Olmos de Le\'on \etal \cite{jmcg:Olmos01}, for instance, fit $\alpha$,  $\beta$ and $\gamma_\pi$.  Unfortunately the dispersion relation approach is not entirely free of assumptions; for instance the asymptotic form of the backward spin-independent amplitude is assumed to be dominated by $\sigma$ exchange.  There is only one attempt, by Baranov \etal \cite{jmcg:BLPS01} to fit the whole data set (up to 150~MeV) which was unfortunately done before the Mainz data was released.

The use of \HBCPT\ with or without the Delta provides an alternative, model-independent approach to the problem, and the history of such attempts is given above.

\section{An analysis of proton Compton data using the delta-expansion}

\begin{figure}
\begin{center}\includegraphics[width=.8\textwidth]{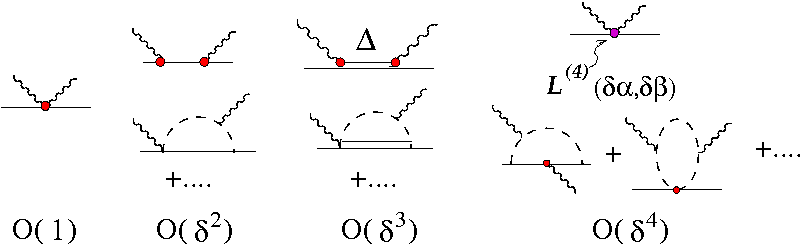}\end{center}
\caption{Sample diagrams that contribute at ${\cal O}(\delta^4)$.}
\label{jmcg:fig1}
\end{figure}

In figure \ref{jmcg:fig1} we show the type of diagrams which contribute at each order.
Most of the parameters which enter are well known.  The values of the second order $\pi N$ scattering LECs $c_i$ which we use must of course be ones appropriate for a Delta-full theory. The $\gamma N\Delta$ coupling constant (variously $b_1$ or $G_M$ in the literature) is not well known; though it has been fit to resonant electroproduction, the power counting in that regime differs from that appropriate to the low-energy regime, and so it is not clear that the same value should be used.  We take the same approach as Hildebrandt \etal \cite{jmcg:HGHP04} and include it as a fit parameter.

There is a problem with the fit when including both $\pi\Delta$ loops and higher-order $\pi N$ terms.  Both individually correct the tendency of the third-order $\pi N$ cross section to be too low in the vicinity of the cusp at backward angles.  Together they over-correct.  It is easy to deduce that the reason  is related to the spin polarisabilities: there is exactly the same over-correction in the value of $\gamma_{M1}$.  We therefore promote a counterterm from higher order to allow us to set this to the best estimates of its experimental value \cite{jmcg:drech99,jmcg:HGHP04} or to fit it to data.

There is an obvious tension in deciding how high to take our energy cut-off in fitting data.  On the one hand the higher we take it, the more data we can include and the lower the statistical error.  On the other hand we know that we are not using the correct power counting for the Delta resonance region and we would not expect a good description of the data there. There are also issues associated with the choice of frame for the calculation which become more acute at higher energies.  We have looked at cut-offs up to 240~MeV lab energy.

\begin{figure}
\includegraphics[width=\textwidth]{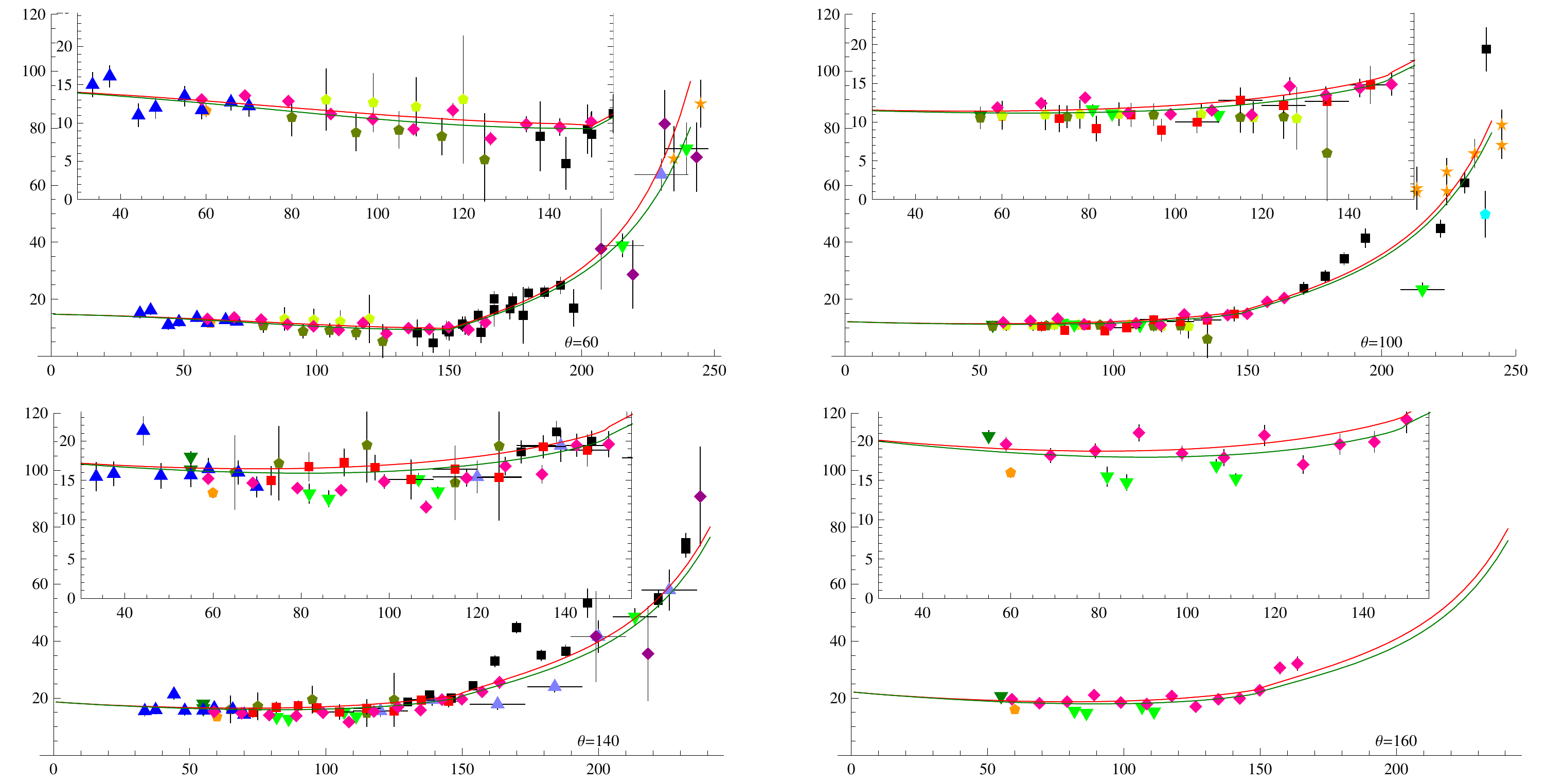}
\caption{Fits to world data: with (green, lower) and without (red, higher) the inclusion of the Delta width and higher order terms in the $\gamma N\Delta$ vertex.  The data are from a wide variety of sources which we will not list here.  For display purposes they have been binned in $40^\circ$ bins.}
\label{jmcg:fig2}
\end{figure}

At this point our fits are preliminary, and we will not quote results.  Figure \ref{jmcg:fig2} however shows a fit demonstrating that we can obtain a good description of data up to at least 200~MeV.  The two curves illustrate the effect of differences in the treatment of the Delta: in one the treatment strictly follows the low-energy power counting, in the other we have included the Delta width and $\Delta/\mn$ corrections to the $\gamma N\Delta$ vertex.  The parameters used are consistent with those obtained by Beane \etal \cite{jmcg:bmmpv03} and Hildebrandt \etal \cite{jmcg:HGHP04}.  In particular a largish value of $\beta$ is preferred, a result consistent with the recent analysis of Lensky and Pascalutsa \cite{jmcg:len09}

A main goal of this study is to produce reliable single-nucleon amplitudes for incorporation in a code for scattering from the deuteron \cite{jmcg:grie09}, which will allow the analysis the data that is currently being taken at MAX-lab \cite{jmcg:feld08}.  We believe we have essentially reached that goal.  However the prospects for reducing the error bars on $\alpha_p$ and $\beta_p$ are not so good.  More high-quality lower-energy data (below pion production) is urgently needed.

\acknowledgments
This research was supported by the US Department of Energy under grants DE-FG02-93ER40756 (DRP) and DE-FG02-95ER-40907(HWG), by the Science and Technology Facilities Council of the UK under grants R105132 (DRP) and R103273 (JMcG), and by the National Science Foundation under career grant PHY-0645498 (HWG and DS).  JMcG acknowledges the hospitality of George Washington University and The Institute for Nuclear Theory, University of Washington, where some of the work was carried out.

\end{document}